\documentclass{elsart}
\usepackage{lineno}
\usepackage{amssymb}
\usepackage{color}
\usepackage{graphicx}
\usepackage{bm}
\def\be{\begin{equation}}
\def\bea{\begin{eqnarray}}
\def\ee{\end{equation}}
\def\eea{\end{eqnarray}}

\def\eps{\varepsilon}

\def\Om{\Omega}
\def\om{\omega}

\def\mod{\mbox{ mod}}

\begin{document}
\begin{frontmatter}

\title{Using Resonances to Control Chaotic Mixing within a Translating and Rotating Droplet}

\author[RC]{R. Chabreyrie}
\author[DV1,DV2]{D. Vainchtein}
\author[CC]{C. Chandre}
\author[PS]{P. Singh}
\author[RC]{N. Aubry}

\address[RC]{Mechanical Engineering Department, Carnegie Mellon University, Pittsburgh, PA 15213, USA}
\address[DV1]{Mechanical Engineering
Department, Temple University, Philadelphia, PA 19122, USA}
\address[DV2]{Space Research Institute, Moscow, GSP-7, 117997, Russia}
\address[CC]{Centre de Physique Th\'eorique\thanksref{CNRS}, CNRS -- Aix-Marseille Universit\'es, Luminy-case 907, F-13288 Marseille cedex 09, France }
\address[PS]{Mechanical Engineering Department, New Jersey Institute of Technology, Newark, NJ 07102, USA}
\thanks[CNRS]{UMR 6207 of the CNRS, Aix-Marseille and Sud Toulon-Var Universities. Affiliated with the CNRS Research Federation FRUMAM (FR 2291). CEA registered research laboratory LRC DSM-06-35.}

\begin{abstract}
Enhancing and controlling chaotic advection or chaotic mixing within liquid droplets is crucial for a variety
of applications including digital microfluidic devices which use microscopic
``discrete'' fluid volumes (droplets) as microreactors. In this work, we
consider the Stokes flow of a translating spherical liquid droplet
which we perturb by imposing a time-periodic rigid-body rotation. Using the
tools of  dynamical systems, we have shown in previous work that the rotation not only
leads to  one or more three-dimensional chaotic mixing regions, in which mixing
occurs through the stretching and folding of material lines, but also offers
the possibility of controlling both the size and the location of  chaotic mixing
within the drop. Such a control was achieved through appropriate tuning of the amplitude and frequency of the rotation in order to use resonances between the natural frequencies of the
system and those of the external forcing. In this paper, we study the influence of the orientation of the rotation axis on the chaotic mixing zones as a third parameter, as well as propose an experimental set up to implement the techniques discussed.  
\end{abstract}


\end{frontmatter}

\section{Introduction\label{introduction}}

The concept of chaotic advection, also referred to as Lagrangian chaos, was introduced some twenty-years ago in order to enhance mixing in laminar flows, the latter being mostly 
two-dimensional, time dependent incompressible flows 
(see \cite{Aref:2002} for a  historical development).\\ 
Several works have shown that the presence of chaotic advection in a flow drastically increases the transport of passive particles. This jump in transport properties has been quantified by the diffusion coefficient 
(see, e.g., the works of \cite{Castiglione:1999, Mathew:2007}; see also  \cite{Solomon:1988a, Solomon:1988b, Paoletti:2006, Mancho:2006} for experimental studies and \cite{Aref:2002, Mancho:2006, Crisanti:1991} for comprehensive reviews).
Chaotic advection is generally obtained by adding a degree of freedom to an incompressible two-dimensional flow.  This degree of freedom can take the form of either time dependence,   \cite{Solomon:1988a, Solomon:1988b, Paoletti:2006, Mancho:2006} or  a third spatial dimension \cite{Baj, KroujilineandStone:1999}.\\
Due to the fact that microfluidic systems are characterized by low Reynolds numbers, typical flows in such devices are laminar and turbulence is inexistent. Thus, one has to turn to a different strategy to achieve mixing, and chaotic advection is often the most efficient way to accomplish this.
 
Microfluidics can either use continuous streams as in the case of microchannels or individual droplets in the so-called ``digital microfluidic devices''.  In the latter, mixing of  multiple reagents takes place within ``discrete'' fluid volumes (droplets), thus offering the possibility of using a multitude of droplets with each droplet playing the role of a microreactor \cite{Ismagilov:2003, Bringer:2004}.\\
Mixing in microfluidic devices using chaotic advection has recently attracted much attention.
While there are many passive strategies based on altering the channel geometry for
flows in microchannels, the use of active techniques based on forcing (see, e.g., \cite{Oddy:2001,
Bau:2001, Ouldelmoctar:2003, GlasgowAubry:2003, Glasgow:2004}) has also proved
to be efficient, especially at low Reynolds numbers \cite{Goullet:2006}. The
combination of both passive and active methods has been explored as well
\cite{Goullet:2006,Niu:2003, Bottausci:2004, Stremler:2004}.\\
Mixing inside a drop subjected to a forcing (at low
Reynolds number) has been studied extensively in the literature \cite{Baj,
KroujilineandStone:1999, Lee:2000, WardandHomsy:2001, Grigoriev:2005,
Homsy:2007, VWG:2007}. It has also been demonstrated experimentally using periodic forcing
\cite{WardandHomsy:2003,GSS:2006}.
 
This article builds upon the work of \cite{ChabreyriePRE08} which investigated the effect of periodic forcing on a
translating droplet and its chaotic fluid flow regions.\\
While previous works \cite{Baj, KroujilineandStone:1999,VVN:1996a,
VNM:2006} have shown the presence of chaotic advection in three-dimensional
bounded steady flows, we concentrate here on unsteady flows and use the added
unsteadiness to manipulate the obtained chaotic behavior through resonances 
\cite{VWG:2007, Lima:1990,  CFP2:1996}.\\
The physical system and the dynamical system which describes it are outlined in Sec.~\ref{dynamical_system}, and the numerical
results are described in Sec.~\ref{numerical_results}. We first recall that the chaotic mixing
zone can be monitored  in both location and size, which we show qualitatively by displaying {\it Liouvillian sections}. We also quantify the size of the mixing zone and study its variation as the amplitude, frequency and orientation of the rotation vary.  In the last section, we propose an experimental device capable of implementing the dynamics studied in this paper.

\section{ Dynamical system \label{dynamical_system}}

\subsection{The dynamical system}

Consider a spherical Newtonian droplet which is itself immersed in an incompressible
Newtonian fluid. The drop undergoes a translating motion as well as a rigid body
rotation, as in the work of \cite{KroujilineandStone:1999}. Furthermore, the droplet is assumed to be spherical and thus the interfacial tension large enough.  We also suppose a very small Reynolds number, i.e. $Re \ll 1$. Given these assumptions, it follows that 
Stokes flow is a reasonable approximation, with the internal and external flows satisfying the boundary
conditions at the surface of the droplet, namely the continuity of velocity and
the tangential stress balance at the interface. In this paper, we further explore the
addition of a degree of freedom to the above problem by making the amplitude of the
rotation periodic in time as in \cite{ChabreyriePRE08}. We also assume that the (mean) amplitude, frequency and orientation of the rotation can be varied.\\
The internal flow is a combination of a steady Hill vortex-type
base flow and a perturbation which takes the form of an oscillating rigid body rotation.
The location $\bm{X}$ of a passive tracer satisfies the dynamical system
\begin{equation}
\bm{\dot{X}}=\bm {V}\left(\bm{X},t\right)= \bm{V}_{0}\left(\bm{X}\right)+ a_{\om}(t)
\hat{\bm{\om}} \times \bm{X},
\label{intflowvect}
\end{equation}
where $\bm{V}$ denotes the velocity of the tracer.
 
In Eq.~(\ref{intflowvect}), the base flow is denoted by $\bm{V}_{0}$ and the perturbation consists of a rotation having a time-periodic amplitude $a_{\om}(t)$ and a
oriented along the unit vector $\hat{\bm{\om}}$.
We now select a moving Cartesian coordinate system translating with the center-of-mass
velocity of the droplet. Let the unit vector $\bm{e}_{z}$ point in the direction
of the translation and the unit vector $\bm{e}_{x}$ lie in the
$\hat{\bm{\om}}-\bm{e}_{z}-$plane. The following non-autonomous
dynamical system then follows:
\begin{equation}
\label{veloT21}
u=\dot{x}= zx - a_{\om}(t) \om_z y , \nonumber 
\end{equation}
\begin{equation}
v=\dot{y}= zy + a_{\om}(t)\left(\om_z x - \om_x z\right),
\label{veloT22} 
\end{equation}
\begin{equation}
\label{veloT23}
w=\dot{z}= 1-2x^2-2y^2-z^2 + a_{\om}(t) \om_x y , 
\end{equation}
where all lengths and velocities have been made dimensionless by normalizing 
with respect to the droplet radius  and the magnitude of the translational velocity. In this work, we assume that the amplitude of the rotation is 
purely sinusoidal about a mean value (it contains only one harmonic), i.e.,
\be
\nonumber
a_{\om}(t)=\frac{\eps}{2}\left(1+\cos \om t \right) ~\mbox{with}~ 0 \le \eps \ll 1,
\label{at}
\ee
and that the orientation of the rotation is given by the unitary vector
\be
\nonumber
\hat{\bm{\om}} =
\left(\om_x,0,\om_z\right)=\left(\cos\alpha,0,\sin\alpha\right).
\label{omega}
\ee
Note that Eq.~ (\ref{intflowvect}) is identical to the equations given in
\cite{KroujilineandStone:1999}, except that the vorticity vector is no longer constant but now given by $a_{\om}(t)\hat{\bm{\om}}$. This
variation in time could result from the presence of unsteady vorticity in the
external flow field or a time dependent body force. In practice, this could be
realized, e.g., by creating a time dependent swirl motion in the external flow
or by applying an electric field capable of exerting a torque on the droplet (this could be achieved by using
traveling wave dielectrophoresis which translates and rotates particles \cite{AubrySingh:2006} or electrorotation which rotates particles \cite{Arn:1988}). Note that this flow is the
superposition of a Hill's vortex and an unsteady rigid body rotation, and that the
surface of the droplet, $r^2 = x^2+y^2+z^2 =1$, is invariant under the flow given by Eqs.
(\ref{veloT21}), (\ref{veloT22}) and (\ref{veloT23}).

\subsection{Base flow}

\begin{figure}
\begin{center}
\includegraphics*[width=14cm]{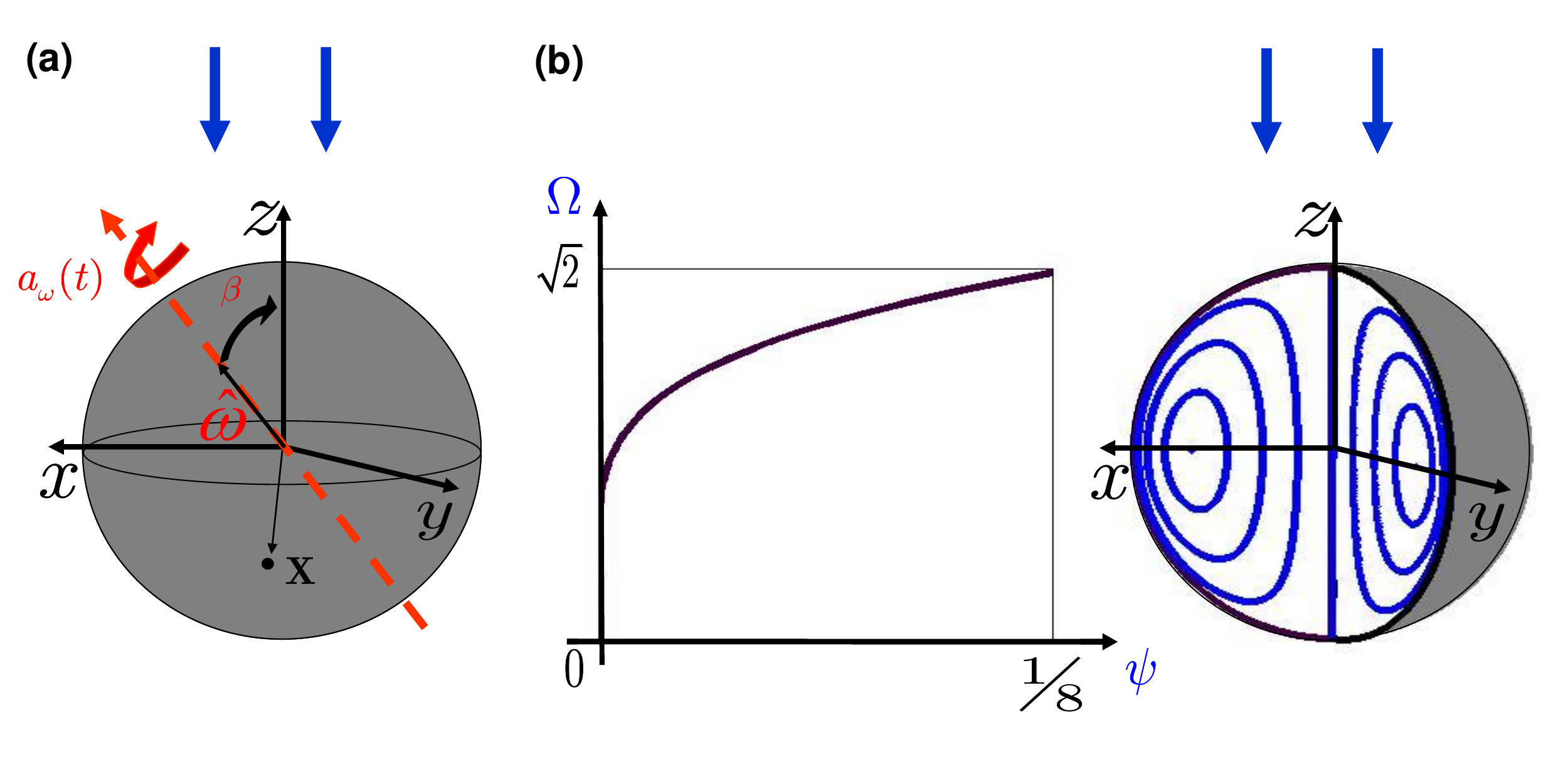}
\caption{(a) Droplet in the perturbed case (with unsteady rigid body rotation);
(b) Streamlines within the droplet for the base (integrable) flow (without rigid body rotation). The motion on each streamline has a frequency $\Om\left(\psi\right)$ given by Eq.~(\ref{Per}).}
\end{center}
\label{figure1_CNSNS}
\end{figure}

In this section, we define the base flow ($\eps=0$), which corresponds to the integrable case in terms of dynamical systems theory. This is a
two-dimensional axisymmetric flow with two independent integrals of
motion, e.g., the streamfunction $\psi$ and the azimuthal angle $\phi$:
\begin{equation}
\nonumber
\psi = \frac12 \left(x^2+y^2 \right)\left(1-r^2\right), \quad \phi =\arctan\left(
y/x\right),
\end{equation}
where $\psi \in \left[0,1/8\right]$. The streamlines, $\Gamma_{\psi,\phi}$,  
are simply lines of constant $\psi$ and $\phi$, satisfying the equation $(1-2x^2-2y^2 )^2+(2 z(x^2+y^2))^2 = 1-8\psi$ (see
Fig.~\ref{figure1_CNSNS}). The surface of the droplet coincides with heteroclinic
orbits defined by $\psi=0$ connecting two hyperbolic fixed points located at the poles of
the spherical drop. As $\psi$ increases from $\psi=0$ to the value $\psi=1/8$, the streamlines are
closed curves converging toward  a circle of degenerate elliptic fixed points
($x^2+y^2=1/2,z=0$). The frequency of the dynamics on a streamline
$\Gamma_{\psi,\phi}$ takes the expression
\begin{equation}
\label{Per}
\frac{2\pi}{\Om(\psi)} = \int^{\pi/2}_{-\pi/2}\frac{\sqrt{2} \;
\beta}{\sqrt{1+\gamma(\psi)\sin\beta}} =\frac{2\sqrt{2}}{\sqrt{1+\gamma}}
K\left( \sqrt{\frac{2\gamma}{1+\gamma}}\right),
\end{equation}
where $\gamma(\psi)=\sqrt{1-8\psi}$ and $K$ refers to the complete elliptic function
of the first kind. The frequency $\Om$ lies in between $\Om(0)=0$ and
$\Om(1/8)=\sqrt{2}$ (see Fig.~\ref{figure1_CNSNS}). We now define a uniform phase $\chi~\mod~(2\pi)$ such that $\chi=0$ on the ${\bm e}_{x}-{\bm e}_{y}-$plane and
$\dot{\chi}=\Om\left(\psi\right)$. Every point in the interior of the droplet, except those lying on the $z-$axis, can be described by the values of $(\psi,\phi,\chi)$. The unperturbed flow can then be expressed in terms of the action-action-angle variables $(\psi,\phi,\chi)$ as follows:
\[
\dot{\psi} =0, \quad \dot{\phi} =0, \quad \dot{\chi} = \Om(\psi).
\]

\subsection{Perturbed flow} 

The perturbed flow is defined by $0<\eps\ll 1$, with the integrals of
motion $\psi$ and $\label{psiphi}$ satisfying
\begin{equation}
\dot{\psi} = -2 a_\om(t)\om_x\psi\sin\phi G\left(\psi,\chi\right),
~~~~\dot{\phi} = a_\om(t)\om_z - a_\om(t)\om_x\cos\phi G\left(\psi,\chi\right),
\label{psiphi} 
\end{equation}
where $G(\psi,\chi)=z/(x^2+y^2)$ is a $2\pi$ periodic function of $\chi$, with zero
average. The time evolution of $\chi$, on the other hand, satisfies the equation 
\[
\dot{\chi}=\Om(\psi)+a_\om(t) H(\psi,\phi, \chi),
\]
where $H$ is a $2\pi$ periodic function of $\chi$. The perturbed system is characterized by two time
scales, a fast one related to the evolution of $\chi$ and a slow one linked to both $\psi$ and $\phi$. As we see below, chaotic advection can be obtained by exploring 
resonance phenomena between the frequency $\Om$ of the unperturbed flow (integral case)  and the forcing frequency $\om$.\\
From a dynamical systems viewpoint, the flow $ \bm{V}(\bm{X},t)$ is the
superimposition of the integrable flow $\bm{V}_{0}(\bm{X})$ and a small time-dependent perturbation $ a_{\om}(t)
\hat{\bm{\om}} \times \bm{X}$, as given in Eq.~(\ref{intflowvect}).
Since the unperturbed flow has two invariants, the trajectories of this
integrable dynamics are all periodic orbits. Most periodic orbits, however, are
expected to break under the influence of a generic time-dependent perturbation with
arbitrarily small amplitude, thus possibly leading to chaotic mixing properties.  
As well-known, the trajectories of an integrable system with only one invariant \cite{FKP:1988} are two-dimensional tori. Perturbing such an integrable system leads to  poor mixing properties since two-dimensional tori (which are robust to small perturbations)
act as barriers to chaotic diffusion. As in our previous work \cite{ChabreyriePRE08}, the goal of this paper is the generation of three-dimensional chaotic mixing regions of a given size and at specific locations. The strategy we adopt consists of bringing a family of
unperturbed tori
$
\left\{  \Gamma_{\psi_n}\right\}_{n \in \mathbb{N}^{*}}
$
 into resonance with the perturbation
$a_{\om}(t)$ 
by adjusting the frequency $\om$ in order to satisfy the resonance
condition:
\begin{equation}
\label{resonance}
n\Om\left(\psi_n\right)=\om.
\end{equation}
The notation $CMR_n$ is then used to denote the chaotic mixing region
created around the torus $ \Gamma_{\psi_n}$.

 We seek to control the mixing by varying the three parameters of the rotation, i.e.,
its amplitude $\eps$, its frequency $\om$ and its orientation $\alpha$. Note that the effects of the amplitude and frequency have been studied for a fixed value of $\alpha$ \cite{ChabreyriePRE08}.

The effect of a rotation is studied, with an amplitude such that $\eps \ll 1$, with $\eps=0$  at which chaotic mixing is inexistent and 
$\eps=\eps_{max}$ at which mixing is maximum.
In addition, we limit our study to the range $0\leq\om\leq\sqrt{2}$ which includes the frequencies $\Om$ of all tori within the droplet.  Notice that  $\Om=0$ on the boundary of the droplet and $\Om=\sqrt{2}$ as the torus approaches the elliptic fixed point. Furthermore, due to the
symmetry of Eqs.~(\ref{veloT21}), (\ref{veloT22}) and (\ref{veloT23})  and without loss of generality, we restrict our study to $0\leq\alpha\leq\pi/2$.

\section{Numerical results \label{numerical_results}}

\subsection{Controlling the location of the chaotic mixing region}

\begin{figure}[htb]
\begin{center}
\includegraphics[width=10cm, height=10cm]{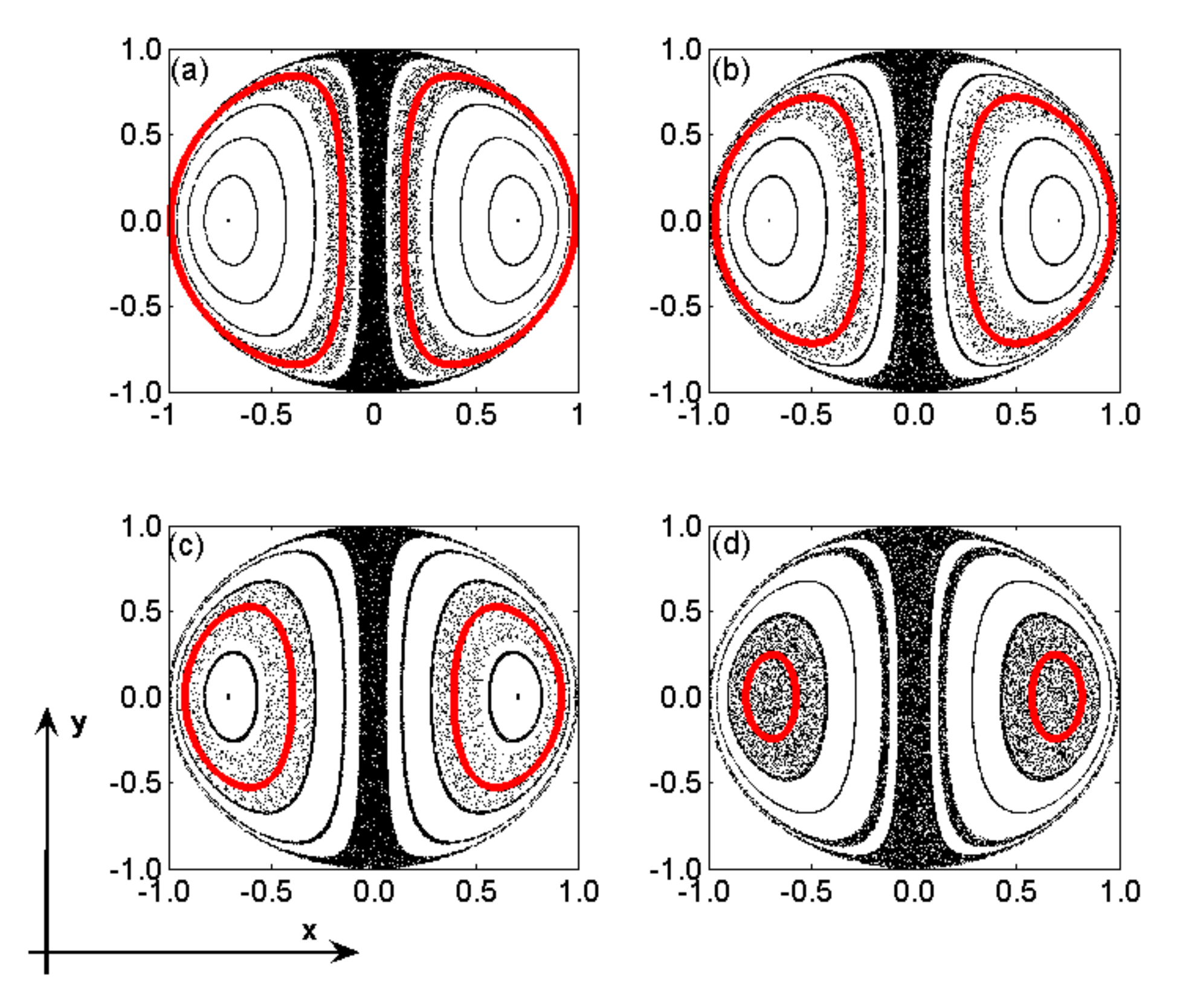}
\caption{Liouvillian sections for the rotation frequencies $\om= 0.95, 1.1, 1.25, 1.40$ (a-d), amplitude $\eps = 0.05$ and orientation $\alpha = \pi/4$. The (red)
line inside the chaotic mixing region $CMR _1$ is the torus $ \Gamma_{\Om^{-1}(\om)}$.}
\end{center}
\label{figure2}
\end{figure}

In this section, we recall our previous findings on the effect of the amplitude and frequency of the rotation for the orientation $\alpha = \pi/4$, as the influence of the orientation $\alpha$ on these results is studied below.\\
Figures~\ref{figure2},~\ref{figure3} and ~\ref{figure4} display the {\it
Liouvillian sections} of the perturbed flow, which consist of two-dimensional projections
of time-periodic three-dimensional flows by a combination of a stroboscopic map
and a plane section. Specifically, the Liouvillian sections considered here are
the intersections of the trajectories with the plane $y=0$ at every period $2\pi/\om$.\\
Figure~\ref{figure2} shows that the perturbation $a_{\om}(t)$ creates two
non-negligible three-dimensional chaotic mixing regions:
one around the torus  having the frequency $\om$ denoted by $CMR_1$ and another one $CMR_{n>1}$ around the pole-to-pole axis 
and near the drop boundary. The latter region contains all tori  having a frequency $\om/n$ with $n>1$.\\
In Fig.~\ref{figure2},  we clearly see that  the location of $CMR_1$  varies with the value 
of $\om$ according to Eq.~(\ref{resonance}). It is important to note that $CMR_{n>1}$ remains around the pole-to-pole axis  and drop boundary
due to the nearly vertical part  of the curve $\Om(\psi)$ close to $\psi=0$ (see Fig.~\ref{figure1_CNSNS}).\\ 
For small values of $\om$, all resonances are located near the pole-to-pole
heteroclinic connections (at $\psi=0$, near the $z-$axis and near the surface
of the droplet, see Fig.~\ref{figure2}). For larger $\om$ values, $CMR_1$
separates from the chaotic region located close to the heteroclinic orbits and
penetrates deeper into the droplet. In the interval $0<\om<
\sqrt{2}$, $CMR_1$ is the largest chaotic region, followed by $CMR_{n>1}$ .
As $\om$ is increased further, $CMR_1$ moves toward the location of the central 
elliptic fixed point by following the location of the
resonant torus with the frequency $\om$.

\subsubsection{Controlling the size of the chaotic mixing region}

\begin{figure}[htb]
\begin{center}
\includegraphics[width=10cm, height=10cm]{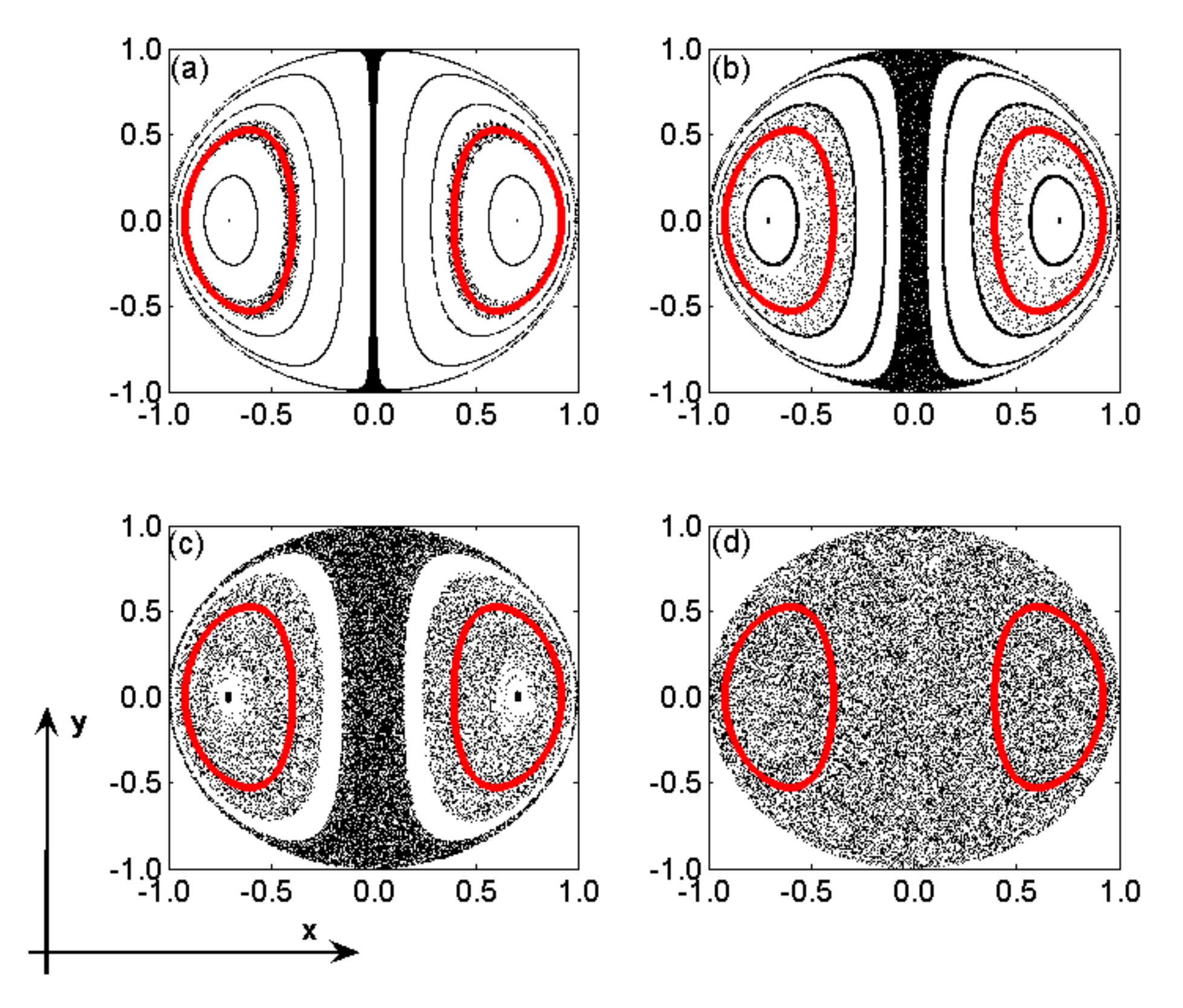}
\caption{(Color on line) Liouvillian sections for
the rotation frequency $\om = 1.25 $, amplitudes $\eps = 0.01, 0.05, 0.10, 0.20$
(a-d) and orientation $\alpha = \pi/4$. The (red) line inside the chaotic mixing region $CMR_1$ is the torus $\Gamma_{\Om^{-1}(\om)}$.}
\end{center}
\label{figure3} 
\end{figure}

In this section, we analyze the size of the two main chaotic mixing regions,
i.e., $CMR_1$ and $CMR_{n>1}$ as the three parameters $\eps$, $\om$ and
$\alpha$ vary. In order to quantify the size, we use the fact that 
for a trajectory starting at $\psi=\psi_0$ the adiabatic
invariant varies between $\psi^-\left(\psi_0; \eps,\alpha,\om \right)$ and
$\psi^+\left(\psi_0;\eps,\alpha,\om \right)$. It then follows that the width
$\Delta \psi= \psi^+\left(\psi_0; \eps,\alpha,\om \right)-\psi^-\left(\psi_0;
\eps,\alpha,\om \right)$ is large close to the resonance but decreases as one goes away from it.  This quantity thus seems to be a
good candidate to quantitatively estimate the
size of $CMR_1$ and $CMR_{n>1}$.\\
As explained above, the location of $CMR_1$ is mostly determined by the rotation frequency $\om$ (while $CMR_{n>1}$ is always located in the neighborhood of the
heteroclinic orbits).   However, the size of $CMR_1$ and $CMR_{n>1}$ can be varied by adjusting the rotation amplitude $\eps$ and orientation $\alpha$. This is illustrated in
Fig.~\ref{figure3} which shows that the size of these chaotic mixing regions
clearly increases with the amplitude of the perturbation. It is
also interesting to note that around $\eps=\eps_{max}\approx 0.20$, the two regions join and invade the entire drop. At that point, complete
chaotic mixing is obtained. The size of $CMR_1$ as a
function of the frequency $\om$ is shown in Fig.~\ref{figure567} (middle
panel). From this figure it is clear that for each value of $\eps$ the size reaches a
maximum for a certain value $\om^m\left(\eps\right)$ of the forcing frequency.

\begin{figure}[htb]
\begin{center}
\includegraphics[width=10cm,height=16cm]{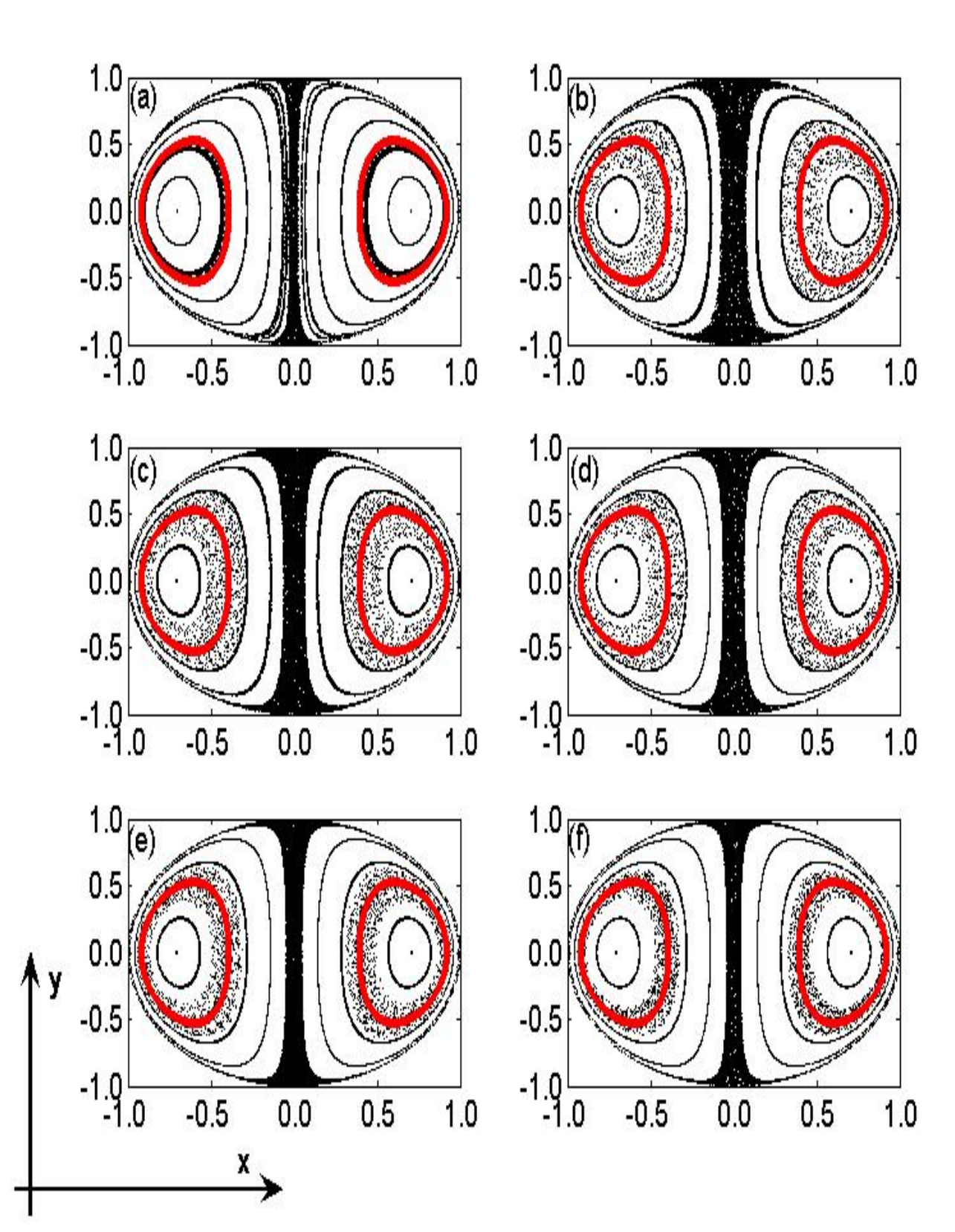}
\caption{Liouvillian sections for the rotation frequency $\om = 1.25 $, amplitude $\eps =
0.05$ and orientations $\alpha =\pi/64, \pi/8, 5\pi/16, 3\pi/8, 7\pi/16$ (a-f).
The (red) line inside the chaotic mixing region $CMR _1$ is the torus $\Gamma_{\Om^{-1}(\om)}$.}
\end{center}
\label{figure4}
\end{figure}

The effect of the orientation $\alpha$ on the size of $CMR_1$ is interesting.  Indeed two distinguishable behaviors can be observed, one in which $\alpha$ hardly affects the size and another one in which $\alpha$ has a strong influence.
These conclusions can be drawn from the lower panel of Fig.~\ref{figure567}, where  the size of the chaotic mixing region $CMR_1$  is displayed as a function of $\alpha$.  
It is indeed clear that when $\alpha$ is not close to the limit values, i.e., $0$ and $\pi/2$, the size stays practically constant, but when $\alpha$ gets close to $0$ and $\pi/2$, it decreases significantly. Such observation is confirmed by looking at the Liouvillian sections in Fig. \ref{figure4}.  In the first two panels, we see a very small change in size while the size of the chaotic mixing regions starts decreasing in the third panel and shrinks drastically in the fourth panel. Being aware of the dependence of the size of $CMR_1$ with respect to the orientation of the rotation $\alpha$ can be handy in practice in order to control the size of the mixing region.\\
Indeed, in some applications where one is faced with the challenge of precisely controlling the size of the mixing zone despite uncontrollable, yet relatively small, fluctuations of $\alpha$, 
setting $\alpha$ far away from the limit values  $0$ and $\pi/2$ and manipulating the chaotic zone size through the parameter $\eps$ should be desirable. 
In other, perhaps more gentle, applications where increasing the amplitude of the rotation may not be possible (e.g., in biomedical handling where 
biological particles need to be handled with care), fixing $\eps$  while tuning the size of the mixing by varying $\alpha$ around the limit values 
$\alpha=0$ or $\alpha=\pi/2$  could be the solution. One should notice, however, that at the critical orientation  $\alpha=\pi/2$ no chaotic  mixing occurs and 
$\psi$  is conserved.\\

\begin{figure}[htb]
\begin{center}
\includegraphics[width=14cm]{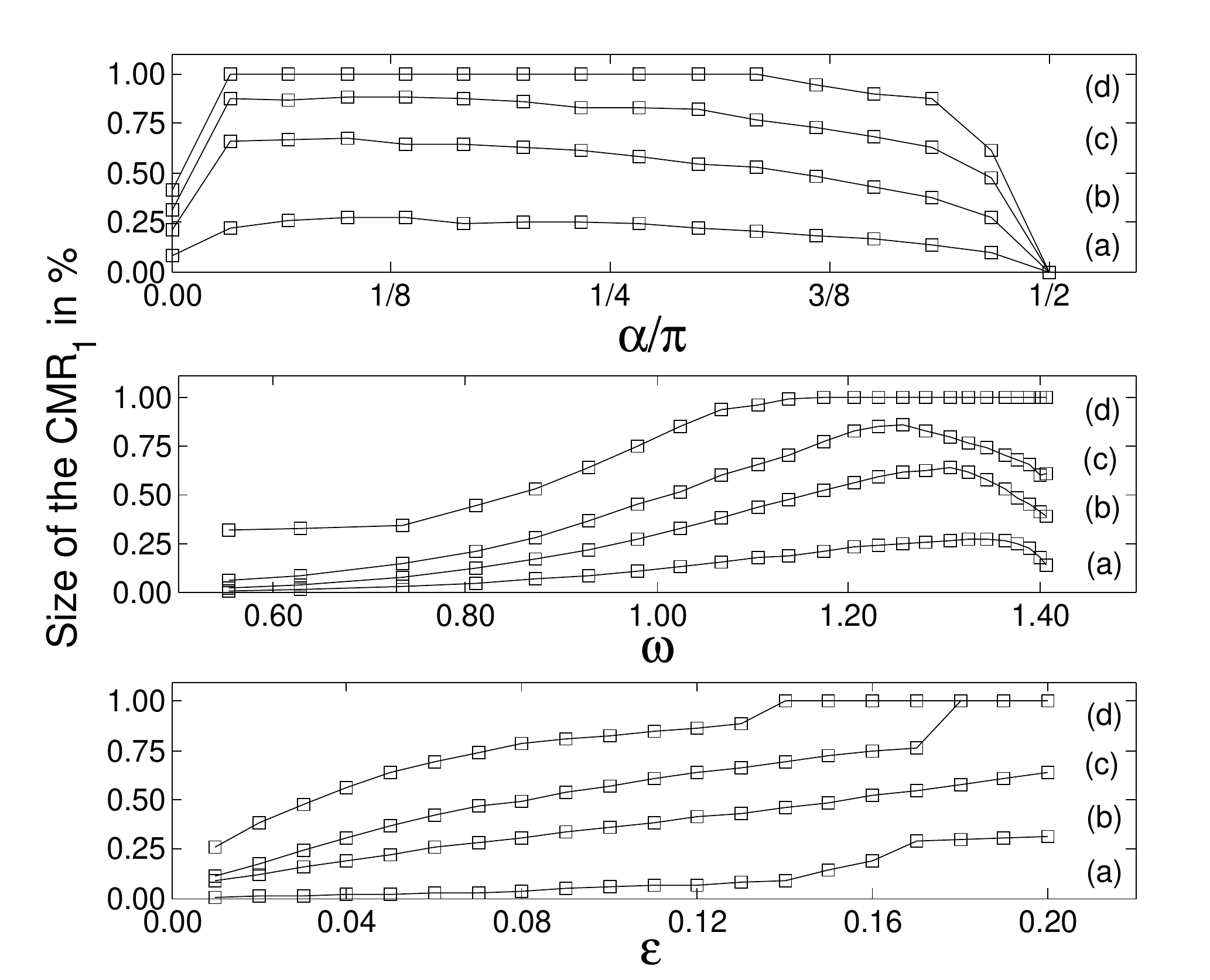}
\caption{Size of the chaotic mixing region.
Upper panel: Normalized size $\Delta\psi$ {\it vs.} orientation
$\alpha$ for the amplitudes $\eps = 0.01, 0.05, 0.10, 0.20$ (a-d) and a  
frequency $\om = 1.25$; Middle panel: Normalized size $\Delta\psi$ {\it vs.} frequency $\om$ for 
the amplitudes $\eps = 0.01, 0.05, 0.10, 0.20$ (a-d) and an orientation $\alpha =
\pi/4$ ; Lower panel: Normalized size $\Delta\psi$ {\it vs.}
amplitude $\eps$ for the frequencies $\om = 0.55, 0.93, 1.28, 1.41$ (a-d) and 
an orientation $\alpha = \pi/4$.}
\end{center}
\label{figure567} 
\end{figure}

\section{\label{sec:expe} Design of an experimental set up}
 
\begin{figure}[t]
\begin{center}
\includegraphics*[width=10cm,height=10cm]{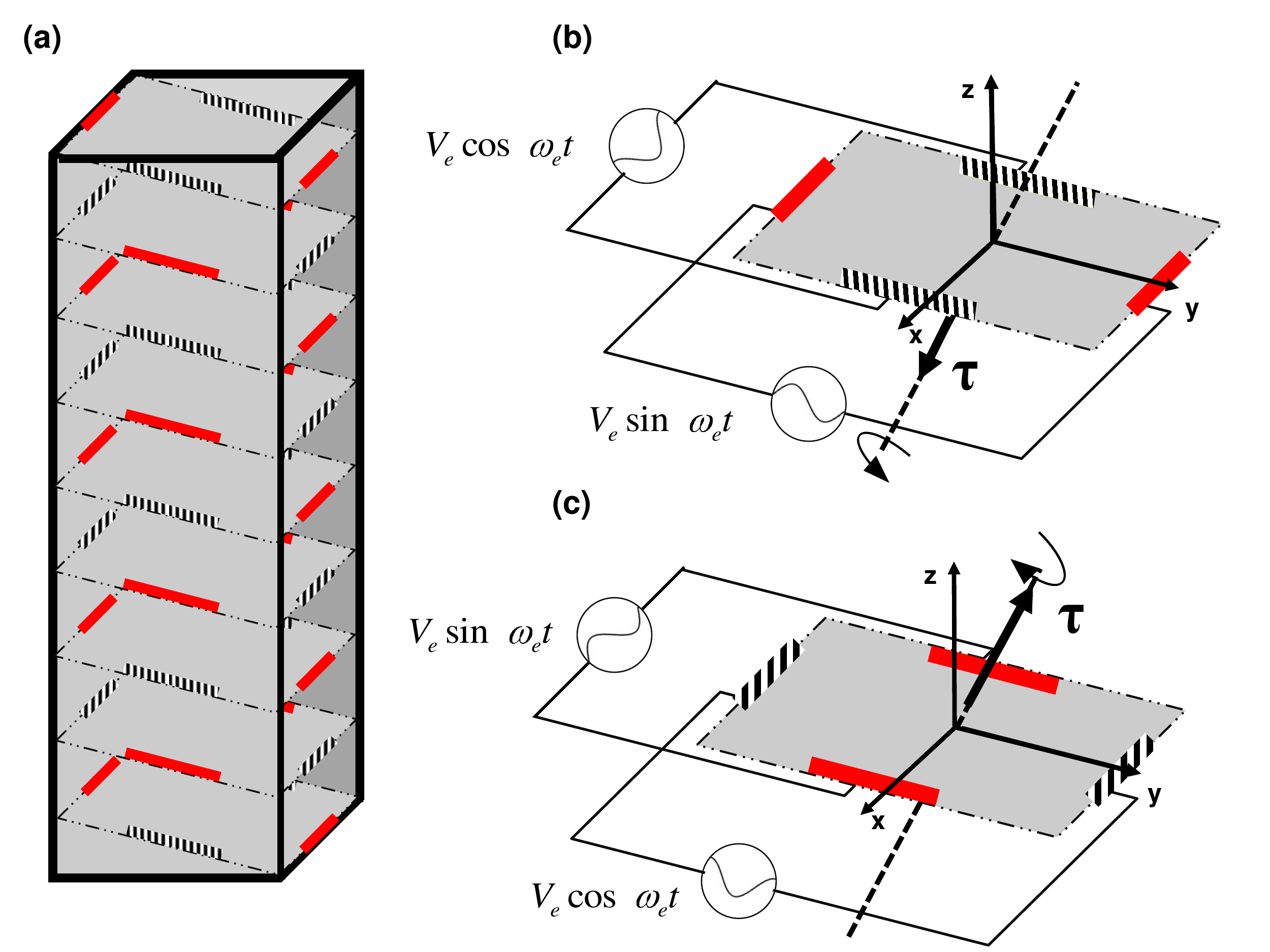}
\end{center}
\caption{Sketch of a possible design for the experimental apparatus showing (a) the square channel/column with a series of out-of-phase embedded electrodes, where $\omega_e$ and $V_e$ stand for the frequency and amplitude of the voltage applied to the electrodes;
(b) the four-pole electrode setting generating a torque in the $z-$direction, i.e., in the direction of translation, $\tau>0$; (c) the four-pole electrode in a consecutive plane setting  generating a torque in the $z-$direction, i.e. in the direction opposed to the direction of translation, $\tau<0$.}
\label{figure7_CNSNS}
\end{figure}

 In this work, the flow within a drop was produced by the superimposition of an external steady translation and an unsteady rigid body rotation.
The drop steady translating motion along the channel could simply be produced by means of a constant pressure gradient, exploring the buoyancy force in the case of a vertical column or using traveling wave dielectrophoresis in a microchannel \cite{AubrySingh:2006}. The unsteady rigid rotation could be realized by applying an electric field that exerts a torque on the droplet as it is the case with traveling wave dielectrophoresis which generates both a force and a torque (although both are not independent) \cite{AubrySingh:2006}.  Another possibility is by using the so-called ``electrorotation'' phenomenon which creates a torque \cite{Arn:1988}. Electrorotation stands for the spinning of an electrically polarized particle   while
the latter is subjected to a ``rotating'' electric field, that is an ac electric field generated by voltages which are out of phase of one another.\\
A possible design based on the latter phenomenon is proposed in Fig.~\ref{figure7_CNSNS}.  This device consists of a vertical square column/channel with electrodes embedded within its four walls creating a rotating electric field due to the phase difference between the voltages applied to adjacent electrodes in the same plane.  It is known that such a four-pole electrode setting would generate a torque on the droplet trapped in the middle of the channel, with the torque strength being proportional to the square of the intensity of the electric field. The periodicity in the angular velocity is then obtained by imposing a phase difference to the voltages applied between two four-pole electrode settings in two different $x-y-$planes $P_{i}$ and $P_{i+1}$, so that a torque around the $z-$direction acts on the drop as it passes through $P_{i}$ and a torque around the $z-$direction acts on the drop as it passes through $P_{i+1}$. Although such a device would not produce exactly the one harmonic angular velocity profile described in Eq. (\ref{at}), our previous study \cite{ChabreyrieMRC08} has shown that the strategy is robust with respect to the specific time dependent function used for the rotation. Specifically, results were very similar when the sinusoidal rotation was replaced by a periodic triangular function. Finally, the control of the orientation of the rotation axis could be realized by a series of electrodes lying within inclined planes along the length of the channel (see Fig.~\ref{figure7_CNSNS}). 
\\

\section{\label{sec:conclu}Conclusions }

In this work, we have further studied the generation of chaotic mixing within a translating droplet by adding a perturbation in the form of an oscillatory rigid body
rotation. As previously \cite{ChabreyriePRE08}, the frequency of the latter was selected in order to create resonances with the natural frequencies of the system, namely the frequencies of the various tori embedded within the drop. A particularly interesting
feature of the perturbed system lies in the fact that both the size and the
location of the mixing region can be varied by adjusting the frequency and amplitude
of the rigid body rotation.\\ 
In this work, we have added a third parameter, namely the orientation of the rotation axis.  It was found that the latter can influence the size of the chaotic mixing regions and that the size can significantly decrease when the angle between the rotation axis and the direction of translation approaches either $0$ or $\pi/2$.  Away from these two limit values, the size of the chaotic mixing region is maximal and the particular value of the angle has only a minor effect.\\ 
A possible design for an experimental set up capable of guiding a drop in a controlled fashion by varying the amplitude, frequency and orientation of the rotation was also proposed.

\section*{ Acknowledgments}

This article is based upon work partially supported by the NSF (grants
CTS-0626070 (N.A.), CTS-0626123 (P.S.) and 0400370 (D.V.)). D.V. is grateful to
the RBRF (grant 06-01-00117) and to the Donors of the ACS Petroleum Research
Fund. C.C. acknowledges support from Euratom-CEA (contract EUR~344-88-1~FUA~F) and CNRS (PICS program).

\end{document}